\documentclass[preprintnumbers,amsmath,amssymb]{revtex4} 

\usepackage{dcolumn}
\usepackage{bm}

\begin{document}

\preprint{UCD 2002-13}
\preprint{DIAS-02-09}

\title{New Scaling Limit for Fuzzy Spheres}

\author{Sachindeo Vaidya}
\email{vaidya@dirac.ucdavis.edu}
\affiliation{Department of Physics, University of California,
Davis, CA 95616, USA.}

\author{Badis Ydri}
\email{ydri@synge.dias.ie}
\affiliation{School of Theoretical Physics,
Dublin Institute for Advanced Studies, Dublin, Ireland.}

\date{\today}

\begin{abstract}
Using a new scaling limit as well as a new cut-off procedure , we
show that $\phi^4$ theory on noncommutative ${\bf R}^4$ can be
obtained from the corresponding theory on fuzzy ${\bf S}^2 \times
{\bf S}^2$. The star-product on this noncommutative ${\bf R}^4$ is
effectively local in the sense that the theory naturally has an ultra-violet
cut-off $\Lambda$ which is inversely proportional to the noncommutativity $\theta$ , i.e $ \Lambda= \frac{2}{\theta}$.  We show that the UV-IR mixing
in this case is absent to one loop in the $2-$point function and also comment on the $4-$point function.
\end{abstract}

\maketitle
\section{Introduction}
Noncommutative field theories provide a rich variety of
interesting conceptual phenomena. The usual focus of research has
been to try to understand the quantum behavior of theories
defined on certain non-compact manifolds like ${\bf R}^{2n}$ or
compact ones like the fuzzy sphere ${\bf S}_F^2$, ${\bf C}{\bf
P}^{n}$ etc . Theories on ${\bf R}^{2n}_{\theta}$ have received
considerable attention not only because they arise in string
theory, but also because of their peculiar formal properties like
UV-IR mixing \cite{mirase} . Theories on compact noncommutative
manifolds possess the attractive feature that they are simply
finite-dimensional matrix models, and thus hold out the hope that
they correspond to regularized versions of quantum field theories
on ordinary manifolds. However, problems like UV-IR mixing are
still present, although in a regularized form which is different
from their noncompact counterparts \cite{vaidya,dolan}. The fuzzy
sphere ${\bf S}_F^2$ can be ``flattened'' (by scaling the radius
$R$ and the cut-off $l$ to infinity) to give the noncommutative
plane, and the UV-IR mixing re-emerges in its usual singular form
\cite{madore} .

Interestingly, as we show in this letter, if we start with a low
energy sector (which we will define more precisely below) of the
theory on ${\bf S}_F^2 \times {\bf S}_F^2$, and simultaneously use
a new scaling to flatten the spheres, we obtain a theory on ${\bf
R}^4_\theta$ that has
ultra-violet cut-off $\Lambda = \frac{2}{\theta}$.  Moreover, this
theory shows no singular UV-IR mixing, as the noncommutative
parameter $\theta$ and the UV cut-off are intimately related.

A popular strategy for studying noncommutative theories is to use
a version of the star-product that is appropriate in near
$\theta=0$ \cite{esgrva}. This $\theta$-expanded theory has many
puzzling features (as emphasized recently in \cite{wulkenhaar}),
which can essentially be traced to the fact that the star-product
has excellent smoothing properties (i.e. it nicely gets rid of
all modes of characteristic length $L << \theta$), whereas the
$\theta$-expanded product does not cut-off high frequency modes.
In this article, we propose a resolution of these puzzles by
defining noncommutative theories on ${\bf R}^2$ or ${\bf R}^4$ as
certain scaling limits of corresponding theories on ${\bf S}_F^2$
and ${\bf S}_F^2 \times {\bf S}_F^2$. In the process, we also
achieve a discretization of noncommutative theories that allows
numerical simulations to study field-theoretic phenomena. Instead
of approximating theories on nc ${\bf R}^2$ or ${\bf R}^4$ by the
usual commutative lattice (but changing the rule for multiplying
functions), a more natural regularization that retains the memory
of underlying spacetime symmetries is to use finite matrix models
on fuzzy sphere ${\bf S}_F^2$ and ${\bf S}_F^2 \times {\bf
S}_F^2$. However , and as we will show , in order that the star
product on ${\bf R}^4$ and ${\bf R}^2$ has the property of
cutting-off top modes of the order of $1/{\theta}$ and higher ,
as it should be , the scaling limit has to be defined precisely.
It is worth noting here that in this re-definition of the Moyal
plane as a sequence of fuzzy spheres , the resulting cut-off is
in fact non-trivial, i.e it is a consequence of the underlying
star product on ${\bf S}^2_F$ and thus it cannot be simply introduced into
the theory . We now explain all this in some detail for the case of ${\bf S}^2_F$ and then in the next sections we derive explicitly all these results for the case of ${\bf S}^2_F{\times}{\bf S}^2_F$ .

The fuzzy sphere is described by three matrices $x_i^F = \theta
L_i$ where $L_i$'s are the generators of $SU(2)$ for the spin $l$
representation and $\theta$ has dimension of length. The radius
$R$ of the sphere is related to $\theta$ and $l$ as $R^2 =
\theta^2 l( l+ 1 )$ . The usual action for a matrix model on
${\bf S}_F^2$ is
\begin{equation}
S = \frac{R^2}{2l+1} {\rm Tr}\; \left( \frac{[L_i, \Phi]^{\dagger}
[L_i , \Phi]}{R^2} + {\mu}_l^2 \Phi^2 + V[\Phi] \right),
\label{fs2action}
\end{equation}
and has the right continuum limit as $l \rightarrow \infty$.
Because of the noncommutative nature of ${\bf S}_F^2$, there is a
natural ultra-violet (UV) cut-off: the maximum energy $\Lambda$
is $= 2l(2l+1)/R^2$. To get the theory on a noncommutative plane,
the usual strategy \cite{madore} is to restrict to the
neighbourhood of (say) the north pole ( where $L_3{\simeq}l$ ) ,
define the noncommutative coordinates as $x_a^{NC}{\equiv}x_a^F$ ,
$a=1,2$, and then take both $l$ and $R$ to infinity with
${\theta}^{'} =\frac{R}{\sqrt{l}}$ fixed, giving us the
commutation relations
\begin{eqnarray}
[{x}_1^{NC}, {x}_2^{NC}]=-i {\theta}^{'2}.\label{0}
\end{eqnarray}
In this limit $\Lambda$ clearly diverges .

Here , we point out another scaling limit in which
$R,l{\longrightarrow}{\infty}$ while keeping $\theta \sim R/l$
fixed . The above noncommutativity relation becomes
simply 
\begin{eqnarray}
[x_1^{NC},x_2^{NC}]=-iR\theta \label{01}
\end{eqnarray}
 which means that $x_a^{NC}$'s
are now strongly noncommuting coordinates ( $R{\rightarrow}{\infty}$ )
and hence nonplanar amplitudes are expected to simply drop out in
accordance with \cite{mirase}. This can also be seen from the fact
that in this scaling (as is obvious from the relation
$R^2={\theta}^2l(l+1)$) $\Lambda$ no longer diverges: it is now
of order $1/\theta$, and there are no momentum modes in the theory
larger then this value. This mode
expansion thus seems appropriate as a starting point for studying
( or even defining perturbatively ) noncommutative field theories.

Alternatively we will also show that in this
limit the noncommutative coordinates can be instead identified as
$X_{a}^{NC}=x_a^{NC}/{\sqrt{l}}$ with noncommutative structure
\begin{equation} 
[{X}_1^{NC}, {X}_2^{NC}]=-i {\theta}^2.\label{1}
\end{equation} 
While this scaling for obtaining ${\mathbb R}^2_{\theta}$ is simply
stated, obtaining the corresponding theory with the above criteria is
somewhat subtle. Indeed passing from $[x_1^{NC},x_2^{NC}]=-iR\theta$
to (\ref{limit2}) corresponds in the quantum theory to a re-scaling of
momenta sending thus the finite cut-off $\Lambda=\frac{2}{\theta}$ to
infinity. In order to bring the cut-off back to a finite value
${\Lambda}_x=x{\Lambda}$, where $x$ is an arbitrary positive real
number, we modify the Laplacian on the fuzzy sphere
${\Delta}=[L_i,[L_i,..]]$ so that to project out modes of momentum
greater than a certain value $j$ given by
$j=[\frac{2\sqrt{l}}{x}]$. In other words, the theory on the
noncommutative plane ${\mathbb R}^2_{\theta}$ with UV cut-off
${\theta}^{-1}$ is obtained by flattening not the full theory on the
fuzzy sphere but only a ``low energy'' sector. One can argue that only
for when ${\Lambda}_x={\Lambda}$ that the canonical UV-IR
singularities become smoothen out. At this value we have
$j=[2\sqrt{l}]$ which marks somehow the boundary between commutative
and noncommutative field theories. This means that this scaling limit as defined above gives a theory on ${\bf
R}^{2}_\theta$ where $\theta$ has now the physical interpretation
of a UV cut-off . 

\section{Fuzzy Scalar Theory}

The generalization to noncommutative ${\bf R}^4$ is obvious: we
work on ${\bf S}_F^2 \times {\bf S}_F^2$ and then take the scaling
limit with $\theta$ fixed, which is the case of most interest to
us in this article. By analogy with (\ref{fs2action}), the scalar
theory with quartic self-interaction on ${\bf S}_F^2 \times {\bf
S}_F^2$ is
\begin{widetext}
\begin{equation}
S = \frac{R_a^2}{2l_a +1}\frac{R_b^2}{2l_b +1}
  Tr_a Tr_b \left( \frac{[L_i^{(a)},
  \Phi]^{\dagger}[L_i^{(a)}, \Phi]}{R_a^2} + \frac{[L_i^{(b)},
  \Phi]^{\dagger}[L_i^{(b)},\Phi]}{R_b^2} + {\mu}_l^2 \Phi^2 +
  \frac{{\lambda}_4}{4!}\Phi^4 \right),
\label{fs2s2act}
\end{equation}
\end{widetext}
where $a$ and $b$ label the first and the second sphere
respectively, and $L_i^{(a,b)}$'s are the generators of rotation
in spin $l_{a,b}$-dimensional representation of $SU(2)$, and
$\Phi$ is a $(2l_a +1)\times (2l_a +1) \otimes (2l_b +1) \times
(2l_b +1)$ hermitian matrix. As $l_a, l_b$ go to infinity, we
recover the scalar theory on an ordinary ${\bf S}^2 \times {\bf
S}^2$. One can argue that it is enough to set $l_a = l_b = l$ and
$R_a = R_b = R$ which corresponds in the limit to a nc ${\bf
R}^4$ with a trivial ${\bf R}^2{\times}{\bf R}^2$ metric. The
general case would only correspond to different deformation
parameters in the two ${\bf R}^2$'s and the extension of all
results is therefore obvious \cite{dn}  .

Following \cite{gkp,vaidya} the fuzzy field ${\Phi}$ can be
expanded in terms of polarization operators \cite{vmk} as follows
\begin{widetext}
\begin{eqnarray}
{\Phi}&=&(2l+1)\sum_{k_1=0}^{2l}\sum_{m_1=-k_1}^{k_1}\sum_{p_1=0}^{2l}\sum_{n_1=-p_1}^{p_1}{\Phi}^{k_1m_1p_1n_1}T_{k_1m_1}(l){\otimes}T_{p_1n_1}(l).\nonumber
\end{eqnarray}
\end{widetext}
Imposing reality , i.e $\hat{\Phi}^{+}=\hat{\Phi}$ , we obtain the
following conditions $
\bar{\phi}^{k_1m_1p_1n_1}=(-1)^{m_1+n_1}{\phi}^{k_1-m_1p_1-n_1}$,
and a canonical path integral quantization will therefore yield
the propagator
\begin{widetext}
\begin{eqnarray}
<{\phi}^{k_1m_1p_1n_1}{\phi}^{k_2m_2p_2n_2}>&=&\frac{(-1)^{m_2+n_2}}{R^2}\frac{{\delta}^{k_1k_2}{\delta}^{m_1,-m_2}{\delta}^{p_1p_2}{\delta}^{n_1,-n_2}}{k_1(k_1+1)+p_1(p_1+1)+R^2{\mu}^2_{l}}.\label{propagator}\nonumber
\end{eqnarray}
\end{widetext}
The Euclidean $4-$momentum in this setting is given by
$11{\equiv}(k_1,m_1,p_1,n_1)$ with square
$(11)^2=k_1(k_1+1)+p_1(p_1+1)$. The vertex is given however by
\begin{eqnarray}
&&S^{int}_{l}
=\sum_{11}\sum_{22}\sum_{33}\sum_{44}V(11,22,33,44){\phi}^{11}{\phi}^{22}{\phi}^{33}{\phi}^{44}\nonumber\\
&&V(11,22,33,44)=R^4\frac{{\lambda}_4}{4!}V_1(1234,km)V_2(1234,pn),\nonumber
\end{eqnarray}
where
\begin{eqnarray}
V_1(1234,km)&=&(2l+1)Tr_{H_{1}}\bigg[T_{k_1m_1}(l)...T_{k_4m_4}(l)\bigg],\nonumber
\end{eqnarray}
and a similar definition for $V_2(1234,pn)$ .

Using standard perturbation theory , the one-loop correction to
the $2-$point function \cite{longpaper} is
\begin{eqnarray}
{\mu}_l^{2}(k_1,p_1)={\mu}_l^{2}+\frac{1}{R^2}\frac{{\lambda}_4}{4!}\bigg[{\delta}{\mu}_l^{P}+{\delta}{\mu}_l^{NP}(k_1,p_1)\bigg]
\end{eqnarray}
with the planar contribution given by
\begin{eqnarray}
&&{\delta}{\mu}_l^{P}=4\sum_{a=0}^{2l}\sum_{b=0}^{2l}A(a,b)~\nonumber\\
&&A(a,b)=\frac{(2a+1)(2b+1)}{a(a+1)+b(b+1)+R^2{\mu}_l^2},\label{planar1}
\end{eqnarray}
whereas the non-planar contribution is  given by
\begin{eqnarray}
{\delta}{\mu}_l^{NP}(k_1,p_1)&=&2
\sum_{a=0}^{2l}\sum_{b=0}^{2l}A(a,b)(-1)^{k_1+p_1+a+b}B_{k_1p_1}(a,b)\nonumber\\
B_{ab}(c,d)&=&(2l+1)^2\{
\begin{array}{ccc}
a&l   & l\\

c& l&l
\end{array}
\}\{
\begin{array}{ccc}
b&l  & l\\

d& l&l
\end{array}
\}.\label{nonplanar1}
\end{eqnarray}
The symbol $\{ \}$ in $B_{ab}(c,d)$ is of course the standard
$6j$ symbol \cite{vmk} . As one can immediately see from the
analytic expressions above , both planar and non-planar graphs
are empty from singularities , everything is finite and well
defined for all finite values of $l$ \cite{denj}. Indeed a measure
for the fuzzy UV-IR mixing or the noncommutative anomaly will be
the
 differences ${\Delta}$
 between  planar and non-planar contributions which can be
 defined by the equation
\begin{eqnarray}
&&{\delta}{\mu}_l^{P}+{\delta}{\mu}_l^{NP}(k_1,p_1)={\Delta}{\mu}_l^P+\frac{1}{2}{\Delta}(k_1,p_1)\nonumber\\
&&{\Delta}{\mu}_l^P=6\sum_{a=0}^{2l}\sum_{b=0}^{2l}A(a,b),\nonumber
\end{eqnarray}
where
\begin{eqnarray}
{\Delta}(k_1,p_1)&=&4\sum_{a=0}^{2l}\sum_{b=0}^{2l}A(a,b)\bigg[(-1)^{k_1+p_1+a+b}B_{k_1p_1}(a,b)-1\bigg]\label{regularized}.
\end{eqnarray}
The fact that this difference is not zero in the limit of
infinite points density , i.e $l{\longrightarrow}{\infty}$ , is what
is meant by UV-IR mixing on fuzzy spaces . (\ref{regularized})
can also be taken as the regularized form of the UV-IR mixing on
${\bf R}^4$  . Removing the UV cut-off
$l{\longrightarrow}{\infty}$ while keeping the infrared cut-off
$R$ fixed $=1$ one can show that ${\Delta}$ diverges as $l^2$ ,
i.e
\begin{eqnarray}
{\Delta}(k_1,p_1)&{\longrightarrow}&(8l^2)\int_{-1}^{1}\int_{-1}^{1}\frac{dt_xdt_y}{2-t_x-t_y}\bigg[P_{k_1}(t_x)P_{p_1}(t_y)-1\bigg],\label{estimation1}
\end{eqnarray}
where , for simplicity , we have assumed ${\mu}_l<<l$
\cite{madore}. (\ref{estimation1}) is worse than the case of two
dimensions [ see equation $(3.20)$ of \cite{madore} ] , in here
not only the difference survives the limit but also it diverges .
This means in particular that the UV-IR mixing can be largely
controlled or perhaps understood if one understands the role of the
UV cut-off $l$ in the scaling limit and its relation to the
underlying star product on ${\bf S}^2_F$.

The computation of higher order correlation functions becomes very
complicated, but for completeness we also write down the result for the $4-$point function . We get
\begin{eqnarray}
{\delta}{\lambda}_4(1235) &=& \frac{{\lambda}_4}{4!} \sum_{k_4,k_6}
\sum_{p_4,p_6} \frac{A(k_4,p_4)
A(k_6,p_6)}{(2k_4+1)(2k_6+1)(2p_4+1)(2p_6+1)} \bigg[8{\eta}_1^{(1)}
{\eta}_2^{(1)} \nonumber\\
&+&16{\eta}_1^{(2)} {\eta}_2^{(2)} + 4{\eta}_1^{(3)} {\eta}_2^{(3)} +
8{\eta}_1^{(4)}{\eta}_2^{(4)}\bigg].
\label{1001}
\end{eqnarray}
The first graph in (\ref{1001}) is the usual one-loop contribution to
the $4-$point function , i.e the two vertices are planar. The fourth
graph contains also two planar vertices but with the exception that
one of these vertices is twisted , i.e with an extra phase. The second
graph contains on the other hand one planar vertex and one non-planar
vertex, whereas the two vertices in the third graph are both
non-planar. The analytic expressions for ${\eta}_i^{(a)} {\equiv}
{\eta}_i^{(a)}(k_4k_6;1235) = \sum_{m_4=-k_4}^{k_4} \sum_{m_6 =
-k_6}^{k_6}{\rho}_i^{(a)}(k_4k_6;1235)$ are given by
\begin{eqnarray}
{\rho}_i^{(1)} &=& (-1)^{m_4 + m_6} V_i(\hat{1} \hat{2}4_f6_f)
V_i(\hat{3} \hat{5}-4_f-6_f), \quad {\rho}_i^{(2)} = (-1)^{m_4 + m_6}
V_i(\hat{1} \hat{2}4_f6_f) V_i(\hat{3}-4_f\hat{5}-6_f) \nonumber\\
{\rho}_i^{(3)} &=& (-1)^{m_4 + m_6} V_i(\hat{1}4_f \hat{2}6_f)
V_i(\hat{3}-4_f \hat{5}-6_f), \quad {\rho}_i^{(4)} = (-1)^{m_4 + m_6}
V_i(\hat{1} \hat{2}4_f6_f) V_i(\hat{3} \hat{5}-6_f-4_f),\nonumber
\end{eqnarray}
where the lower index in $\eta$'s and $\rho$'s labels the sphere
whereas the upper index denotes the graph, and the notation $-4_f4_f$
stands for $(k_4,-m_4,p_4,-n_4)$ in contrast with
$4_f4_f = (k_4,m_4,p_4,n_4)$.

By using extensively the different identities in \cite{vmk} we can
find after a long calculation that the above $4$-point function has
the form
\begin{eqnarray}
{\delta}{\lambda}_4(1235) &=& \frac{{\lambda}_4}{4!} \bigg[ 8 {\delta}
{\lambda}_4^{(1)}(1235) + 16 {\delta}{\lambda}_4^{(2)}(1235) + 4
{\delta}{\lambda}_4^{(3)}(1235) + 8 {\delta}{\lambda}_4^{(4)}(1235)
\bigg], \quad \rm{where} \nonumber\\
{\delta}{\lambda}_4^{(a)}(1235) &=& \sum_{k_4,k_6} \sum_{p_4,p_6}
A(k_4,p_4) A(k_6,p_6) {\nu}^{(a)}_1(k_4k_6;1235)
{\nu}^{(a)}_2(p_4p_6;1235), \;\;a=1 \ldots 4,
\label{1002}
\end{eqnarray}
The label $f$ stands for the shells we integrated over and hence it
corresponds to $q^2=(2l+1)^2$ for the full one-loop contribution. The
planar amplitudes, in the first ${\mathbb R}^2$ factor for example,
are given by
\begin{equation} 
{\nu}_1^{(1)} = \sum_{k}(-1)^{k + k_4 + k_6} {\delta}_{k}(1235)
E_{k_1k_2}^{k_4k_6}(k) E_{k_3k_5}^{k_4k_6}(k), \quad {\nu}_1^{(4)} =
\sum_{k}{\delta}_{k}(1235) E_{k_1k_2}^{k_4k_6}(k)
E_{k_3k_5}^{k_4k_6}(k)
\label{planar4}
\end{equation} 
whereas the non-planar amplitudes are given by
\begin{equation} 
{\nu}_1^{(2)} = \sum_{k}(-1)^{k_3 + k_4} {\delta}_{k}(1235)
E_{k_1k_2}^{k_4k_6}(k) F_{k_3k_5}^{k_4k_6}(k), \quad {\nu}_1^{(3)} =
\sum_{k}(-1)^{k_2 + k_3}{\delta}_{k}(1235) F_{k_1k_2}^{k_6k_4}(k)
F_{k_3k_5}^{k_4k_6}(k)
\label{nonplanar4}
\end{equation} 
with
\begin{eqnarray}
F_{k_1k_2}^{k_4k_6}(k)&=&(2l+1)\sqrt{(2k_1+1)(2k_2+1)} \left\{
\begin{array}{ccc}
k_4&l  & l\\
k_6&l&l\\
k&k_1&k_2
\end{array}
\right\}\nonumber\\
E_{k_1k_2}^{k_4k_6}(k)&=&(2l+1)\sqrt{(2k_1+1)(2k_2+1)}\left\{
\begin{array}{ccc}
k_1&k_2  & k\\
l&l&l
\end{array}
\right\} \left\{
\begin{array}{ccc}
k_4&k_6  & k\\
l&l&l
\end{array}
\right\}.
\label{F}
\end{eqnarray}
The ``fuzzy delta'' function $\delta_k(1235)$ is defined by
\begin{equation} 
{\delta}_{k}(1235) = (-1)^m C^{km}_{k_1m_1k_2m_2}
C^{k-m}_{k_3m_3k_5m_5}.
\label{delta}
\end{equation}

\section{Continuum Planar Limit}

We can now state with some detail the continuum limits in which the
fuzzy spheres approach (in a precise sense) the noncommutative
planes. There are primarily two limits of interest to us: one is the
canonical large stereographic projection of the spheres onto planes,
while the second is a new flattening limit which we will argue
corresponds to a conventional cut-off.

For simplicity, consider a single fuzzy sphere with cut-off $l$ and
radius $R$, and define the fuzzy coordinates $x_i^{F}={\theta}L_i^{}$
(i.e.  $x_{\pm}^{F}=x_1^{F}{\pm}ix_2^{F}$) where
${\theta}=R/{\sqrt{l(l+1)}}$. The stereographic projection
onto the noncommutative plane is realized as
\begin{equation} 
y_{+}^{F}=2R^{}x_{+}^{F}\frac{1}{R-x_3^{F}}, \quad
y_{-}^{F}=2R\frac{1}{R-x_3^{F}}x_{-}^{F}. 
\end{equation} 
In the large $l$ limit it is obvious that these fuzzy coordinates
indeed approach the canonical stereographic coordinates. A planar
limit can be defined from above as follows:
\begin{equation} 
{\theta}^{'2}=\frac{{R}^{2}}{\sqrt{l(l+1)}}=\text{fixed as}\quad 
l,R{\rightarrow}{\infty}.
\label{contrast}
\end{equation} 
In this limit, the commutation relation becomes
\begin{equation} 
[y_{+}^{NC},y_{-}^{NC}] = -2{\theta}^{'2}, \quad
y_{\pm}^{NC}{\equiv}y_{\pm}^{F} = x_{\pm}^F,
\end{equation} 
where we have substituted $L_{3}=-l$ corresponding to the north
pole. The above commutation relation may also be put in the form
\begin{equation} 
[x^{NC}_1,x^{NC}_2] = -i{\theta}^{'2}, \quad x^{NC}_a{\equiv}x^{F}_a,
\quad a=1,2
\label{comm1}
\end{equation} 
The minus sign is simply due to our convention for the coherent states
on co-adjoint orbits. The extension to the case of two fuzzy spheres
is trivial.

A second way to obtain the noncommutative plane is by taking the limit
\begin{equation} 
{\theta}=\frac{R}{\sqrt{l(l+1)}}=\quad l,R{\rightarrow}{\infty}.
\label{flattening2}
\end{equation}
A UV cut-off is automatically built into this limit: the maximum
energy a scalar mode can have on the fuzzy sphere is $2l(2l+1)/R^2$,
which in this scaling limit is $4/\theta^2$. There are no modes with
energy larger than this value. To understand this limit a little
better, let us restrict ourselves to the north pole
$\vec{n}=\vec{n}_0=(0,0,1)$ where we have $\langle
\vec{n}_0,l|L_3|\vec{n}_0,l \rangle = -l$ and $\langle
\vec{n}_0,l|L_a|\vec{n}_0,l \rangle=0$, $a=1,2$. The commutator
$[L_1,L_2]=iL_3 = -il$, so the noncommutative coordinates on this
noncommutative plane ``tangential to the north pole'' can be given
either simply by $x_a^F$ as above. This now defines a strongly
noncommuting plane, viz
\begin{eqnarray}
[x_a^F,x_b^F]=-il{\theta}^2{\epsilon}_{ab}.\label{or}
\end{eqnarray}
Or aletrnatively one can define the noncommutative coordinate by 
${X}^{NC}_a{\equiv}\sqrt{\frac{\theta}{R}}x_a^F$, satisfying
\begin{equation}
[X^{NC}_a,X^{NC}_b]=-i{\theta}^2{\epsilon}_{ab}.
\label{commutationrelation}
\end{equation}
In the convention used here, ${\epsilon}_{12}=1$ and
${\epsilon}_{ac}{\epsilon}_{cb} = -{\delta}_{ab}$.

Intuitively, the second scaling limit may be understood as
follows. Noncommutativity introduces a short distance cut-off of the
order ${\delta}X=\sqrt{\frac{{\theta}^2}{2}}$ because of the
uncertainty relation $ {\Delta}X^{NC}_1 {\Delta}X^{NC}_2 {\geq}
\frac{{\theta}^2}{2}$. However, the Laplacian operators on generic
noncommutative planes do not reflect this short distance cut-off, as
they are generally taken to be the same as the commutative
Laplacians. On the above noncommutative plane
(\ref{commutationrelation}) the cut-off ${\delta}X$ effectively
translates into the momentum space as some cut-off ${\delta}P =
\frac{1}{\sqrt{2{\theta}^2}}$. This is because of (and in accordance
with) the commutation relations $[X^{NC}_a, P^{NC}_b] = i
\delta_{ab}$,$P_a^{NC} =
-\frac{1}{{\theta}^2}{\epsilon}_{ab}X_b^{NC}$, giving us the
uncertainty relations ${\Delta}X^{NC}_a {\Delta}P^{NC}_b {\geq}
\frac{{\delta}_{ab}}{2}$. Since one can not probe distances less than
${\delta}X$, energies above ${\delta}P$ should not be accessible
either, i.e.  $[P_a^{NC},P_b^{NC}] =
-\frac{i}{{\theta}^2}{\epsilon}_{ab}$. The fact that the maximum
energy of a mode is of order $1/\theta$ in the second scaling limit
ties in nicely with this expectation.

The limit (\ref{flattening2}) may thus be thought of as a
regularization prescription of the noncommutative plane which takes
into account our expectation of ``UV-finiteness'' of noncommutative
quantum field theories.

\subsection{Field Theory in the Canonical Planar Limit}

We are now in a position to study what happens to the scalar field
theory in the limit (\ref{comm1}). First we match the spectrum of the
Laplacian operator on each sphere with the spectrum of the Laplacian
operator on the limiting noncommutative plane as follows
\begin{equation} 
a(a+1)=R^2p_{a}^2,
\label{matchingcond0}
\end{equation} 
where $p_a$ is of course the modulus of the two dimensional momentum
on the noncommutative plane which corresponds to the integer $a$, and
has the correct mass dimension. However since the range of $a$'s is
from $0$ to $2l$, the range of $p_a^2$ will be from $0$ to $
\frac{2l(2l+1)}{R^2}=l{\Lambda}^{'2}{\rightarrow}\infty$,
${\Lambda}^{'}=2/{\theta}^{'}$. In other words, all information about
the UV cut-off is lost in this limit.

Let us see how the other operators in the theory scales in the above
planar limit. It is not difficult to show that the free action scales
as
\begin{equation} 
\sum_{a,b} \sum_{m_a,m_b} \bigg[R^2a(a+1)+R^2b(b+1)+R^4{\mu}_l^2\bigg]
|{\phi}^{abm_am_b}|^2 {\simeq} \int_{\sqrt{l}{\Lambda}^{'}}
\frac{d^2\vec{p}_ad^2\vec{p}_b}{{\pi}^2} \bigg[p_a^2+p_b^2+M^2\bigg]
|{\phi}_{NC}^{p_ap_b{\phi}_a{\phi}_b}|^2.
\label{dimensionanalysis0}
\end{equation} 
The scalar field is assumed to have the scaling property
${\phi}^{p_ap_b{\phi}_a{\phi}_b}_{NC}{\simeq} R^4{\phi}^{abm_am_b}$,
which gives the momentum-space scalar field the correct mass dimension
of $-3$ [recall that $[{\phi}^{abm_am_b}]=M$]. The ${\phi}_a$ and
${\phi}_b$ above (not to be confused with the scalar field!) are the
angles of the two momenta $\vec{p}_a$ and $\vec{p}_b$ respectively,
i.e. ${\phi}_a = \frac{{\pi}m_a}{a+\frac{1}{2}}$ and ${\phi}_b =
\frac{{\pi}m_b}{b+\frac{1}{2}}$. This formula is exact, and can be
simplified further when quantum numbers $a$'s and $b$'s are large: the
${\phi}_a$ and ${\phi}_b$ will be in the range $[-{\pi},{\pi}]$. It is
also worth pointing out that the mass parameter $M$ of the planar
theory is exactly equal to that on the fuzzy spheres,
i.e. $M={\mu}_{\infty}$, and no scaling is required.

With these ingredients, it is not then difficult to see that the
flattening limit of the planar $2-$point function (\ref{planar1})
is given by
\begin{equation} 
\delta M^P {\equiv}\frac{{\delta}{\mu}_l^P}{R^2}= 16 \int \int
\frac{p_ap_bdp_adp_b}{p_a^2+p_b^2 + M^2}
\end{equation} 
which is the $2$-point function on noncommutative ${\mathbb R}^4$ with
a Euclidean metric ${\mathbb R}^2 \times {\mathbb R}^2$. By rotational
invariance it may be rewritten as
\begin{equation} 
{\delta} M^P = \frac{4}{{\pi}^2} \int_{\sqrt{l}{\Lambda}^{'}} \frac{d^4k}{k^2+M^2}.
\label{planar10}
\end{equation} 
We do now the same exercise for the non-planar $2$-point function
(\ref{nonplanar1}). Since the external momenta $k_1$ and $p_1$ are
generally very small compared to $l$ , one can use the following
approximation for the $6j$-symbols \cite{vmk}
\begin{eqnarray} 
\left\{\begin{array}{ccc}
           a&l  & l\\
           b& l&l
       \end{array} \right\}{\approx} \frac{(-1)^{a+b}}{2l} P_{a}
           (1-\frac{b^2}{2l^2}), \quad
           l{\rightarrow}{\infty},\;\;a<<l,\;0 {\leq} b{\leq}2l,
\label{appr1}
\end{eqnarray} 
By putting in all the ingredients of the planar limit we obtain the
result
\begin{eqnarray} 
{\delta}M^{NP}(k_1,p_1){\equiv}\frac{{\delta}{\mu}_l^{NP}}{R^2} = 8
\int_{0}^{\infty} \int_{0}^{\infty} 
\frac{p_ap_bdp_adp_b}{p_a^2 + p_b^2 + M^2} P_{k_1}(1 -
\frac{{\theta'}^4 p_a^2}{2R^2}) P_{p_1}(1 -
\frac{{\theta'}^4 p_b^2}{2R^2}). \nonumber
\end{eqnarray} 
Although the quantum numbers $k_1$ and $p_1$ in this limit are very
small compared to $l$, they are large themselves
i.e. $1<<k_1,p_1<<l$. On the other hand, the angles ${\nu}_{a}$
defined by $\cos{\nu}_{a}=1-\frac{{\theta'}^4 p_a^2}{2R^2}$ can be
considered for all practical purposes small, i.e. ${\nu}_a =
\frac{{\theta'}^2 p_a}{R}$ because of the large $R$ factor, and hence
we can use the formula (see for eg \cite{magnus}, page $72$)
\begin{eqnarray} 
P_{n}(\cos{\nu}_a) = J_{0}(\eta) + \sin^2\frac{{\nu}_a}{2}
\bigg[\frac{J_{1}(\eta)}{2{\eta}} - J_{2}(\eta) + \frac{\eta}{6}
J_{3}(\eta)\bigg] + O(\sin^4\frac{{\nu}_a}{2}),
\label{formula}
\end{eqnarray} 
for $n>>1$ and small angles ${\nu}_a$, with $\eta = (2n+1) \sin
\frac{{\nu}_a}{2}$. To leading order we then have
\begin{eqnarray} 
P_{k_1}(1 - \frac{{\theta'}^4 p_a^2}{2R^2}) =
J_0({\theta'}^2 p_{k_1}p_a) = \frac{1}{2{\pi}} \int_{0}^{2{\pi}}
d{\phi}_ae^{i{\theta'}^2 \cos{\phi}_a p_{k_1}p_a}.\nonumber
\end{eqnarray}
This result becomes exact in the strict limit of $l,R \rightarrow
\infty$ where all fuzzy quantum numbers diverge with ${R}$. We get
then
\begin{eqnarray} 
\delta M^{NP}(p_{k_1},p_{p_1}) = \frac{2}{{\pi}^2} \int
\int \int \int
\frac{(p_adp_ad{\phi}_a) (p_bdp_bd{\phi}_b)}{p_a^2+p_b^2 +
M^2}e^{i{\theta'}^2 p_{k_1}(p_a
\cos{\phi}_a)}e^{i{\theta'}^2 p_{p_1}(p_b \cos{\phi}_b)}.\nonumber
\end{eqnarray} 
By rotational invariance we can set ${\theta'}^2 B^{{\mu}{\nu}}
p_{k_1\mu} p_{a\nu} = {\theta'}^2 p_{k_1} (p_a \cos{\phi}_a)$, where
$B^{12}=-1$. In other words, we can always choose the two-dimensional
momentum $p_{k_1}$ to lie in the $y$-direction, thus making ${\phi}_a$
the angle between $\vec{p}_a$ and the $x$-axis. The same is also true
for the other exponential. We thus obtain the canonical non-planar
$2$-point function on the noncommutative ${\mathbb R}^4$ (with
Euclidean metric ${\mathbb R}^2 \times {\mathbb R}^2$). Again by
rotational invariance, this non-planar contribution to the $2$-point
function may be put in the compact form
\begin{eqnarray} 
\delta M^{NP}(p) = \frac{2}{{\pi}^2}\int_{\sqrt{l}{\Lambda}^{'}} \frac{d^4k}{k^2 + M^2}
e^{i{\theta'}^2 p B k}.
\label{nonplanar10}
\end{eqnarray}
The structure of the effective action in momentum space allows us to
deduce the star products on the underlying noncommutative space. For
example, by using the tree level action (\ref{dimensionanalysis0})
together with the one-loop contributions (\ref{planar10}) and
(\ref{nonplanar10}) one can find that the effective action obtained in
the large stereographic limit (\ref{contrast}) is given by
\begin{eqnarray} 
\int_{\sqrt{l}{\Lambda}^{'}} \frac{d^4\vec{p}}{(2{\pi})^4}\frac{1}{2} \bigg[\vec{p}^2 + M^2 +
\frac{g_4^2}{6} \big[2\int_{\sqrt{l}{\Lambda}^{'}} \frac{d^4k}{(2{\pi})^4} \frac{1}{k^2 + M^2}
+ \int_{\sqrt{l}{\Lambda}^{'}} \frac{d^4k}{(2{\pi})^4} \frac{e^{i{\theta}^{'2} \vec{p}B
\vec{k}}}{k^2 + M^2}\big]\bigg]|{\phi}_{1}(\vec{p})|^2
\label{effective0}
\end{eqnarray} 
where $g^2_4=8{\pi}^2{\lambda}_4$ and ${\phi}_{1}(\vec{p}) =
4{\pi}\sqrt{2}{\phi}_{NC}^{p_ap_b{\phi}_a{\phi}_b}$ and
$\sqrt{l}{\Lambda}^{'}{\rightarrow}{\infty}$. This effective 
action can be obtained from the quantization of the action
\begin{eqnarray} 
\int d^4x \bigg[\frac{1}{2}({\partial}_{\mu}{\phi}_1)^2 +
\frac{1}{2}M^2{\phi}_1^2 + \frac{g_4^2}{4!} {\phi}_1 *'
{\phi}_1 *' {\phi}_1 *' {\phi}_1\bigg],\nonumber
\end{eqnarray} 
where ${\phi}_1{\equiv}{\phi}_1(x^{NC})=\int
\frac{d^4p}{(2{\pi})^4}{\phi}_{1}(\vec{p})e^{-ipx^{NC}}={\phi}_1^{\dagger}$
and $*'$ is the canonical (or Moyal-Weyl) star product
\begin{equation} 
f *' g(x^{NC}) = e^{\frac{i}{2}{\theta}^{'2} B^{{\mu}{\nu}}
{\partial}_{\mu}^y {\partial}_{\nu}^z} f(y)g(z)|_{y=z=x^{NC}}
\label{star}
\end{equation} 
This is consistent with the commutation relation (\ref{comm1}) and
provides a nice check that that the canonical star product on the
sphere derived in \cite{dolan} (also given here by equation
(\ref{early})) reduces in the limit (\ref{contrast}) to the above
Moyal-Weyl product (\ref{star}). In the above, $B$ is the
antisymmetric tensor which can always be rotated such that the non
vanishing components are given by $B^{12}=-B_{21}=-1$ and
$B^{34}=-B_{43}=-1$.
 
In fact one can read immediately from the above effective action that
the planar contribution is quadratically divergent as it should be,
i.e.
\begin{eqnarray}
{\Delta}M^P=\frac{1}{64{\pi}^2}{\delta}M^P =
\int_{\sqrt{l}{\Lambda}^{'}{\rightarrow}{\infty}}
\frac{d^4k}{(2{\pi})^4}\frac{1}{k^2+M^2} =
\frac{1}{16{\pi}^2}l{\Lambda}^{'2}{\rightarrow}{\infty}, 
\end{eqnarray}
whereas the non-planar contribution is clearly finite 
\begin{eqnarray} 
{\Delta}M^{NP}(p)=\frac{1}{32{\pi}^2}{\delta}M^{NP}(p) &=&
\int_{\sqrt{l}{\Lambda}^{'}{\rightarrow}{\infty}}
\frac{d^4k}{(2{\pi})^4}\frac{e^{i{\theta}^{'2}\vec{p}B\vec{k}}}{k^2+M^2}
\nonumber\\
&=&\frac{1}{8{\pi}^2}\bigg[\frac{2}{E^2{{\theta'}^4}} + M^2
\ln({\theta'}^2 EM)\bigg], \quad {\rm where} \quad
E^{\nu}=B^{{\mu}{\nu}}P_{\mu}.
\end{eqnarray} 
This is the answer of \cite{mirase}: it is singular at $P=0$ as well
as at $\theta'=0$.

\subsection{A New Planar Limit With Strong Noncommutativity}
As explained earlier, the limit (\ref{flattening2}) possesses the
attractive feature that a momentum space cut-off is naturally built
into it. In addition to obtaining a noncommutative plane in the strict
limit, UV-IR mixing is completely absent. But while the new scaling is
simply stated, obtaining the corresponding field theory is somewhat
subtle. We will need to modify the Laplacian on the fuzzy sphere to
project our modes with momentum greater than $2\sqrt{l}$. In other
words, the noncommutative theory on a plane with UV cut-off $\theta$
is obtained not by flattening the full theory on the fuzzy sphere, but
only a ``low energy'' sector, corresponding to momenta upto $2
\sqrt{l}$.

In order to clarify the chain of arguments, we will first implement
naively the limit (\ref{flattening2}) and show that it corresponds to
a strongly noncommuting plane . Finite noncommuting plane is only
obtainable if we pick a specific low energy sector of the fuzzy sphere
before taking the limit as we will explain in the next section.

Our rule for matching the spectrum on the fuzzy sphere with that on
the noncommutative plane is the same as before, namely
$a(a+1)=R^2p_{a}^2$. However because of (\ref{flattening2}), the range
of $p_a^2$ is now from $0$ to $\frac{2l(2l+1)}{R^2} =
\frac{4}{{\theta}^2}$. The kinetic part of the action will scale in
the same way as in (\ref{dimensionanalysis0}), only now the momenta
$\vec{p}$'s in (\ref{dimensionanalysis0}) are restricted such that
$p{\leq}{\Lambda}$. With this scaling information, we can see that the
planar contribution to the $2$-point function is given by
\begin{equation} 
{\delta}m^P{\equiv}\frac{{\delta}{\mu}_l^{P}}{R^2} = \frac{4}{{\pi}^2} \int_{k{\leq}{\Lambda}}
\frac{d^4k}{k^2 + {\mu}^2_l}, \quad {\Lambda} = \frac{2}{\theta}.
\label{planar20}
\end{equation} 
We can similarly compute the non-planar contribution to the $2$-point
function using (\ref{appr1}). The motivation for using this
approximation is more involved and can be explained as follows. In the
planar limit $l,R{\rightarrow}{\infty}$, it is obvious that the
relevant quantum numbers $k_1$ and $p_1$ are in fact much larger
compared to $1$, i.e. $k_1{\sim}R p_{k_1}>>1$ and $p_1{\sim}R
p_{p_1}>>1$, since $R{\simeq}{\theta}l$. However (\ref{appr1}) can be
used only if $k_1,p_1<<l$, or equivalently $\frac{k_1}{l} =
\frac{2p_{k_1}}{\Lambda}<<1$ and $\frac{p_1}{l} = \frac{2
p_{p_1}}{\Lambda}<<1$. This is clearly true for small external momenta
$p_{k_1}$ and $p_{p_1}$, which is exactly the regime of interest in
order to see if there is UV-IR mixing. The condition for the
reliability of the approximation (\ref{appr1}) is then ${\theta}
p_{external}<<1$. We will sometimes refer to this condition as
``$\theta$ small'', the precise meaning of this phrase being
``momentum scale of interest is much smaller than $1/\theta$''. We
thus obtain
\begin{equation} 
{\delta}m^{NP}(k_1,p_1){\equiv}\frac{{\delta}{\mu}_l^{NP}}{R^2} =
8\int_{0}^{\Lambda} \int_{0}^{\Lambda} 
\frac{p_ap_bdp_adp_b}{p_a^2 + p_b^2 + {\mu}^2_l} P_{k_1}(1 -
\frac{{\theta}^2 p_a^2}{2}) P_{p_1}(1 - \frac{{\theta}^2
p_b^2}{2}).
\end{equation} 
Now the angles ${\nu}_a$'s of (\ref{formula}) are defined by
$\cos{\nu}_{a}=1-\frac{{\theta}^2p_a^2}{2}$, and since ${\theta}p<<1$,
these angles are still small. They are therefore given to the leading
order in $\theta p$ by ${\nu}_a={\theta}p_a+ \cdots$ where the
ellipsis indicate terms third order and higher in $\theta p$. By using
(\ref{formula}) we again have
\begin{equation} 
P_{Rp_{k_1}}(1 - \frac{{\theta}^2 p_a^2}{2}) = J_0(R {\theta}
p_{k_1}p_a) = \frac{1}{2{\pi}} \int_{0}^{2{\pi}} d{\phi}_a
e^{i{R}{\theta} \cos{\phi}_a p_{k_1} p_a}.
\label{well}
\end{equation} 
Using rotational invariance we can rewrite this as
\begin{equation} 
{\delta}m^{NP}(p) = \frac{2}{{\pi}^2} \int_{k{\leq}{\Lambda}}
\frac{d^4k}{k^2 + {\mu}^2_l} e^{iR{\theta}pBk}.
\label{nonplanar20}
\end{equation} 
One immediate central remark is in order: the noncommutative phase
contains now a factor $R{\theta}$ instead of the naively expected
factor of ${\theta}^2$. This is in contrast with the previous case of
canonical planar limit, where the strength of the noncommutativity
${\theta'}^2$ defined by the commutation relation (\ref{comm1}) is
exactly what appears in the noncommutative phase of
(\ref{nonplanar10}). In other words this naive implementation of
(\ref{flattening2}) yields in fact the strongly noncommuting plane
(\ref{or}) instead of (\ref{commutationrelation}). Also we can
similarly to the previous case put together the tree level action
(\ref{dimensionanalysis0}) with the one-loop contributions
(\ref{planar20}) and (\ref{nonplanar20}) to obtain the effective
action
\begin{eqnarray} 
&&\int_{{\Lambda}} \frac{d^4\vec{p}}{(2{\pi})^4} \frac{1}{2}
\bigg[\vec{p}^2 + {\mu}^2_l + \frac{g_4^2}{6} \big[2 \int_{{\Lambda}}
\frac{d^4k}{(2{\pi})^4} \frac{1}{k^2 + {\mu}^2_l} + \int_{{\Lambda}}
\frac{d^4k}{(2{\pi})^4} \frac{1}{k^2 + {\mu}^2_l} e^{iR{\theta} \vec{p} B
\vec{k}} \big] \bigg]|{\phi}_{3}(\vec{p})|^2 .
\label{effective}
\end{eqnarray} 
As before $g_4^2=8{\pi}^2{\lambda}_4$, whereas
${\phi}_3(\vec{p})=l^{3/2}{\phi}_2(\sqrt{l}\vec{p})$,
${\phi}_{2}(\vec{p}) = {4{\pi}}\sqrt{\frac{2}{l^3}}
{\phi}_{NC}(\frac{\vec{p}}{\sqrt{l}})$ with ${\phi}_{NC}(\vec{p})
{\equiv} {\phi}_{NC}^{p_ap_b{\phi}_a{\phi}_b} = R^4{\phi}^{abm_am_b}$
(in the metric ${\mathbb R}^2 \times {\mathbb R}^2$). It is not
difficult to see that the one-loop contributions ${\delta}m^P$ and
${\delta}m^{NP}(p)$ given in (\ref{planar20}) and (\ref{nonplanar20})
can also be given by the equations
\begin{eqnarray}
\bar{\Delta}m^{P} &=& \frac{l}{64{\pi}^2}{\delta}m^P =
\int_{\sqrt{l}{\Lambda}{\rightarrow}{\infty}}\frac{d^4k}{(2{\pi})^4}
\frac{1}{k^2 + l{\mu}^2_l}\nonumber\\ 
\bar{\Delta}m^{NP}(p) &=&
\frac{l}{32{\pi}^2}{\delta}m^{NP}(\frac{p}{\sqrt{l}}) =
\int_{\sqrt{l}{\Lambda}{\rightarrow}{\infty}} \frac{d^4k}{(2{\pi})^4}
\frac{1}{k^2 + l{\mu}^2_l} e^{i{\theta}^2 \vec{p} B \vec{k}}.
\end{eqnarray}
We have already computed that the leading terms in $\bar{\Delta}m^{P}$
and $\bar{\Delta}m^{NP}(p)$ are given by
\begin{eqnarray}
\bar{\Delta}m^{P} = \frac{l}{16{\pi}^2}\bigg[{\Lambda}^2 - {\mu}^2_l
  \ln(1+\frac{{\Lambda}^2}{{\mu}^2_l}\bigg],\;\;\bar{\Delta}m^{NP}(p) 
=\frac{1}{8{\pi}^2}\bigg[\frac{2}{E^2{{\theta}^4}} + l{\mu}^2_l 
\ln({\theta}^2 \sqrt{l}E{\mu}_l)\bigg], \quad {\rm where} \quad
E^{\nu}=B^{{\mu}{\nu}}p_{\mu}.\nonumber
\end{eqnarray}
Obviously then we obtain
\begin{equation} 
{\delta}m^P = 4\bigg[{\Lambda}^2 -
  {\mu}^2_lln(1+\frac{{\Lambda}^2}{{\mu}^2_l}\bigg], \quad
  {\delta}m^{NP}(p) = 4{\mu}^2_l \ln(l{\theta}^2E{\mu}_l).
\end{equation} 
If we now require the mass ${\mu}_l$ in (\ref{dimensionanalysis0}) to
scale as ${\mu}_l^2 = \frac{m^2}{l}$ (the reason will be clear
shortly), then one can deduce immediately that the planar contribution
${\delta}m^P$ is exactly finite equal to $4{\Lambda}^2$, whereas the
non-planar contribution ${\delta}m^{NP}(p)$ vanishes in the limit
$l{\rightarrow}{\infty}$.

Remark finally that despite the presence of the cut-off ${\Lambda}$ in
the effective action (\ref{effective}), this effective action can
still be obtained from quantizing
\begin{equation} 
\int d^4x \bigg[\frac{1}{2} ({\partial}_{\mu} {\phi}_3)^2 +
\frac{1}{2}{\mu}_l^2{\phi}_3^2 + \frac{g_4^2}{4!} {\phi}_3* {\phi}_3*
{\phi}_3* {\phi}_3\bigg],
\label{ncr4action}
\end{equation} 
only we have to regularize all integrals in the quantum theory with a
cut-off $\Lambda=2/\theta$. [${\phi}_3{\equiv}{\phi}_3(x^{F}) = \int
\frac{d^4p}{(2{\pi})^4} {\phi}_{3} (\vec{p}) e^{-ipx^{F}} =
{\phi}_3^{\dagger}$, and the star product $*$ is the Moyal-Weyl
product given in (\ref{star}) with the obvious substitution
${\theta}'{\rightarrow}R{\theta}$].

\subsection{A New Planar Limit With Finite Noncommutativity}

Neverthless, the action (\ref{effective}) can also be understood in
some way as the effective action on the noncommutative plane
(\ref{commutationrelation}) with finite noncommutativity equal to
${\theta}^2$. Indeed by performing the rescaling
$\vec{p}{\rightarrow}\frac{\vec{p}}{\sqrt{l}}$ we get
\begin{eqnarray} 
&&\int_{\sqrt{l}{\Lambda}} \frac{d^4\vec{p}}{(2{\pi})^4} \frac{1}{2}
\bigg[\vec{p}^2 + m^2 + \frac{g_4^2}{6} \big[2 \int_{\sqrt{l}{\Lambda}}
\frac{d^4k}{(2{\pi})^4} \frac{1}{k^2 + m^2} + \int_{\sqrt{l}{\Lambda}}
\frac{d^4k}{(2{\pi})^4} \frac{1}{k^2 + m^2} e^{i{\theta}^2 \vec{p} B
\vec{k}} \big] \bigg]|{\phi}_{2}(\vec{p})|^2 .
\label{effective1}
\end{eqnarray} 
We have already the correct noncommutativity $\theta^2$ in the phase
and the only thing which needs a new reintepretation is the fact that
the cut-off is actually given by
$\sqrt{l}{\Lambda}{\rightarrow}{\infty}$ and not by the finite cut-off
$\Lambda$. [Remark that if we do not reduce the cut-off
$\sqrt{l}{\Lambda}$ again to the finite value ${\Lambda}$, the physics
of (\ref{effective1}) is then essentially that of canonical
noncommutativity, i.e the limit (\ref{flattening2}) together with the
above rescaling of momenta is equivalent to the limit
(\ref{contrast})].

Now having isolated the $l$-dependence in the range of momentum space
integrals in the effective action (\ref{effective1}), we can argue
that it is not possible to get rid of this $l$-dependence merely by
changing variables. Actually, to correctly reproduce the theory on the
noncommutative ${\mathbb R}^4$ given by (\ref{flattening2}) and
(\ref{commutationrelation}), we will now show that one must start with
a modified Laplacian (or alternately propagator) on the fuzzy space
\cite{denj}. For this, we replace the Laplacian ${\Delta} =
[L_i^{(a)},[L_i^{(a)},..]]$ on each fuzzy
sphere which has the canonical obvious spectrum $k(k+1)$,
$k=0,...,2l$, with the modified Laplacian
\begin{eqnarray} 
{\Delta}_j = {\Delta} + \frac{1}{\epsilon}(1 - P_j).
\label{presc}
\end{eqnarray} 
Here $P_j$ is the projector on all the modes associated with the
eigenvalues $k=0,...,j$, i.e.
\begin{eqnarray} 
P_j = \sum_{k=0}^{j}\sum_{m=-k}^k|k,m \rangle \langle k,m|, \nonumber
\end{eqnarray} 
The integer $j$ thus acts as an intermediate scale, and using the
modified propagator gives us a low energy sector of the full
theroy. We will fix the integer $j$ shortly.

With this modified Laplacian, modes with momenta larger than $j$ do
not propagate: as a result, they make no contribution in momentum sums
that appear in internal loops. In other words, summations like
$\sum_{0}^{2l}$ (which go over to integrals with range
$\int_0^\Lambda$) now collapse to $\sum_{0}^{j}$ (where the integrals
now are of the range $\int_0^{{\Lambda}_j}$, with ${\Lambda}_j =
\frac{j}{2l} \Lambda$).

The new flattening limit is now defined as follows: start with the
theory on $S_F^2 \times S_F^2$, but with the modified propagator
(\ref{presc}). First take $\epsilon \rightarrow 0$, then $R,l
\rightarrow \infty$ with $\theta = R/l$ fixed. This gives us the
effective action (\ref{effective1}) but with with momentum space
cut-off $\sqrt{l}{\Lambda}_j=\frac{j}{2\sqrt{l}} \Lambda$ ,i.e
\begin{eqnarray} 
&&\int_{\sqrt{l}{\Lambda}_j} \frac{d^4\vec{p}}{(2{\pi})^4} \frac{1}{2}
\bigg[\vec{p}^2 + m^2 + \frac{g_4^2}{6} \big[2 \int_{\sqrt{l}{\Lambda}_j}
\frac{d^4k}{(2{\pi})^4} \frac{1}{k^2 + m^2} + \int_{\sqrt{l}{\Lambda}_j}
\frac{d^4k}{(2{\pi})^4} \frac{1}{k^2 + m^2} e^{i{\theta}^2 \vec{p} B
\vec{k}} \big] \bigg]|{\phi}_{2}(\vec{p})|^2 
\label{effective2}
\end{eqnarray} 
This also tells us that the correct choice of the intermediate scale
is $j =[2 \sqrt{l}]$ for which $\sqrt{l}{\Lambda}_j={\Lambda}$. For
this value of the intermediate cut-off, we obtain the noncommutative
${\mathbb R}^4$ given by (\ref{flattening2}) and
(\ref{commutationrelation}) .

By looking at the product of two functions of the fuzzy sphere, we can
understand better the role of the intermediate scale $j
(=[2\sqrt{l}])$. The fuzzy spherical harmonics $T_{l_a m_a}$ go over
to the usual spherical harmonics $Y_{l_a m_a}$ in the limit of large
$l$, and so does their product, provided their momenta are
fixed. Alternately, the product of two fuzzy spherical harmonics $T$'s
is ``almost commutative'' (i.e. almost the same as that of the
corresponding $Y$'s) if their angular momentum is small compared to
the maximum angular momentum $l$, whereas it is ``strongly
noncommutative'' (i.e. far from the commutative regime) if their
angular momenta are sufficiently large and comparable to $l$. The
intermediate cut-off tells us precisely where the product goes from
one situation to the other: Working with fields having momenta much
less than $[2\sqrt{l}]$ leaves us in the approximately commutative
regime, while fields with momenta much larger than $[2\sqrt{l}]$ take
us in the strongly noncommutative regime. In other words, the
intermediate cut-off tells us where commutativity and noncommutativity
are in delicate balance. Indeed by writing (\ref{effective2}) in the
form

\begin{eqnarray} 
&&\int_{\sqrt{l}{\Lambda}_j} \frac{d^4\vec{p}}{(2{\pi})^4} \frac{1}{2}
\bigg[\vec{p}^2 + m^2 + \frac{g_4^2}{6} \big[2 \int_{\sqrt{l}{\Lambda}_j}
\frac{d^4k}{(2{\pi})^4} \frac{1}{k^2 + m^2} + \int_{\sqrt{l}{\Lambda}_j}
\frac{d^4k}{(2{\pi})^4} \frac{1}{k^2 + m^2} e^{i{\theta}^2 \vec{p} B
\vec{k}} \big] \bigg]|{\phi}_{2}(\vec{p})|^2{\equiv}\nonumber\\
&&\int_{\Lambda} \frac{d^4\vec{p}}{(2{\pi})^4} \frac{1}{2}
\bigg[\vec{p}^2 + {\mu}^2_{l,j}+ \frac{g_4^2}{6} \big[2 \int_{{\Lambda}}
\frac{d^4k}{(2{\pi})^4} \frac{1}{k^2 + {\mu}^2_{l,j}} + \int_{{\Lambda}}
\frac{d^4k}{(2{\pi})^4} \frac{1}{k^2 + {\mu}^2_{l,j}} e^{i(\frac{j}{2\sqrt{l}})^2{\theta}^2 \vec{p} B
\vec{k}} \big] \bigg]|{\phi}^{(j)}_{3}(\vec{p})|^2.\nonumber
\label{effective3}
\end{eqnarray}
[${\mu}^2_{l,j}=l{\mu}^2_l(\frac{2\sqrt{l}}{j})^2$ ,
${\phi}^{(j)}_3(\vec{p})=(\frac{j}{2\sqrt{l}})^3{\phi}_2(\frac{j}{2\sqrt{l}}\vec{p})$
, ${\phi}^{(2l)}_3{\equiv}{\phi}_3$]. For $j<<[2\sqrt{l}]$,
$(\frac{j}{2\sqrt{l}})^2{\theta}^2{\rightarrow}0$ and this is the
effective action on a commutative ${\mathbb R}^4$ with cut-off
${\Lambda}=2/\theta$. For $j>>[2\sqrt{l}]$ this effective action
corresponds to canonical noncommutativity if we insist on the first
line above as our effective action or to strongly noncommuting
${\mathbb R}^4$ if we consider instead the effective action to be
given by the second line. For the value $j=[2\sqrt{l}]$, where we
obtain the noncommutative ${\mathbb R}^4$ given by (\ref{flattening2})
and (\ref{commutationrelation}), there seems to be a balance between
the above two situations and one can also expect the UV-IR mixing to
be smoothen out.

To show this we write first the one-loop planar and non-planar
contributions for $j=[2\sqrt{l}]$ , viz
\begin{eqnarray}
{\Delta}m^P&=&\int_{{\Lambda}}
\frac{d^4k}{(2{\pi})^4} \frac{1}{k^2 + m^2}~,~
{\Delta}m^{NP}(p)=\int_{{\Lambda}}
\frac{d^4k}{(2{\pi})^4} \frac{1}{k^2 + m^2} e^{i{\theta}^2 \vec{p} B
\vec{k}}. \nonumber
\end{eqnarray}
We can evaluate these integrals by introducing a Schwinger parameter
$(k^2+m^2)^{-1}=\int
d{\alpha}exp\bigg(-\alpha(k^2+m^2)\bigg)$. Explicitly, we obtain for
the planar contribution
\begin{eqnarray}
{\Delta}m^P &=& \frac{1}{16{\pi}^2}\bigg[ -{\Lambda}^2\int
  \frac{d\alpha}{\alpha}e^{-\alpha(m^2+{\Lambda}^2)} + \int
  \frac{d\alpha}{{\alpha}^2}e^{ -\alpha m^2}\bigg(1 - e^{-\alpha
  {\Lambda}^2}\bigg)\bigg] \nonumber\\
&=&\frac{1}{16{\pi}^2}\bigg[{\Lambda}^2+m^2 \ln\frac{m^2}{m^2+{\Lambda}^2}\bigg].
\end{eqnarray}

Obviously the above planar function diverges quadratically as
${\Lambda}^2$ when $\theta{\rightarrow}0$, i.e. the noncommutativity
acts effectively as a cut-off.

Next we compute the non-planar integral. To this end we introduce as
above a Schwinger parameter and rewrite the integral as follows

\begin{eqnarray}
{\Delta}m^{NP}(p)&=&
\frac{1}{16{\pi}^4}\int_{0}^{\infty}
d{\alpha}e^{-{\alpha}m^2-\frac{{\theta}^4E^2}
  {4{\alpha}}}\int_{{\Lambda}}d^4k e^{-{\alpha}\big[\vec{k} -
    \frac{i{\theta}^2}{2{\alpha}}\vec{E}\big]^2}\nonumber\\ 
&=&\frac{1}{16{\pi}^4} \sum_{r=0}^{\infty}({\theta}^2)^r
\sum_{s=0}^{[\frac{r}{2}]}\frac{i^{r-s}}{s!(r-2s)!}\bigg[\int_{0}^{\infty} 
d{\alpha}(\frac{E^2}{4i\alpha})^se^{-{\alpha}m^2-\frac{{\theta}^4E^2}
{4{\alpha}}}\bigg[\int_{{\Lambda}}d^4k e^{-\alpha
k^2}(\vec{k}\vec{E})^{r-2s}\bigg]\bigg]~,~E^{\nu}=B^{\mu
\nu}p^{\mu}.\nonumber
\end{eqnarray}
In above we have also used the fact that $\theta$ is small in the
sense we explained earlier (i.e. $E\theta<<1$) and in accordance with
\cite{wulkenhaar} to expand the second exponential around
$\theta=0$. This is also because the cut-off ${\Lambda}$ is inversely
proportional to $\theta$. [In the last line we used the identity
$\sum_{p=0}^{\infty}\sum_{q=0}^pA_{q,p-q}=\sum_{r=0}^{\infty}\sum_{s=0}^{[\frac{r}{2}]}A_{s,r-2s}$,
$[\frac{r}{2}]=\frac{r}{2}$ for $r$ even and
$[\frac{r}{2}]=\frac{r-1}{2}$ for $r$ odd] . It is not difficult to
argue that the inner integral above vanishes unless $r$ is even. Using
also the fact that the cut-off ${\Lambda}$ is rotationally invariant
one can evaluate the inner integral as follows. We have
\begin{eqnarray}
\int_{{\Lambda}}d^4ke^{-\alpha k^2}(\vec{k}\vec{E})^{n}
&=& 4{\pi}^2E^n(n-1)!!\bigg[ \frac{1}{(2\alpha)^{\frac{n}{2} + 2}} -
  {\Lambda}^ne^{-\alpha{\Lambda}^2}
  \sum_{q=-1}^{\frac{n}{2}}\frac{1}{(n-2q)!!}
  \frac{1}{(2\alpha)^{q+2}}\frac{1}{{\Lambda}^{2q}}\bigg],\nonumber 
\end{eqnarray}
where $n$ is an even number given by $n=r-2s$. 

We can now put the above non-planar function in the form
\begin{eqnarray}
&&{\Delta}m^{NP}(p)
=\frac{1}{16{\pi}^2}\sum_{N=0}^{\infty}\frac{1}{N!}\big(\frac{{\theta}^2E}{2}\big)^{2N}\int
\frac{d\alpha}{{\alpha}^{N+2}}e^{-\alpha
m^2-\frac{{\theta}^4E^2}{4\alpha}}\sum_{M=0}^{N}C_N^M(-1)^M\bigg[1-\sum_{P=0}^{M+1}\frac{(\alpha
{\Lambda}^2)^P}{P!}e^{-\alpha
{\Lambda}^2}\bigg].\label{uvir}\nonumber\\
\end{eqnarray}
[$C_N^M=\frac{N!}{M!(N-M)!}$]. The first term in this expansion
corresponds exactly to the case of canonical noncommutativity where
instead of ${\Lambda}$ we have no cut-off, i.e.
\begin{eqnarray}
{\Delta}m^{NP}(p)
&=&\frac{1}{16{\pi}^2}\sum_{N=0}^{\infty}\frac{1}{N!}\big(\frac{{\theta}^2E}{2}\big)^{2N}\int
\frac{d\alpha}{{\alpha}^{N+2}}e^{-\alpha
m^2-\frac{{\theta}^4E^2}{4\alpha}}\sum_{M=0}^{N}C_N^M(-1)^M+...\nonumber\\
&=& \frac{1}{8{\pi}^2}\bigg[\frac{2}{{\theta}^4E^2} + m^2
  \ln(m{\theta}^2E)\bigg] +...{\equiv} \frac{1}{16{\pi}^2}I^{(2)}(m^2,
\frac{{\theta}^4E^2}{4})+...\nonumber
\end{eqnarray}
As expected this term provides essentially the canonical UV-IR
mixing. As it turns out this singular behaviour is completely
regularized by the remaining $N=0$ term in (\ref{uvir}), i.e.
\begin{eqnarray}
{\Delta}m^{NP}(p)&=&\frac{1}{16{\pi}^2}I^{(2)}(m^2,\frac{{\theta}^4E^2}{4})
+\frac{1}{16{\pi}^2}\int \frac{d\alpha}{{\alpha}^{2}}e^{-\alpha
m^2-\frac{{\theta}^4E^2}{4\alpha}}\bigg[-\sum_{P=0}^{1}\frac{(\alpha
{\Lambda}^2)^P}{P!}e^{-\alpha {\Lambda}^2}\bigg]+......\nonumber\\
&=& \frac{1}{16{\pi}^2} I^{(2)}(m^2,\frac{{\theta}^4E^2}{4}) -
\frac{1}{16{\pi}^2} \bigg[I^{(2)}(m^2 +
  {\Lambda}^2,\frac{{\theta}^4E^2}{4}) +
  {\Lambda}^2I^{(1)}(m^2+{\Lambda}^2, \frac{{\theta}^4E^2}{4})\bigg] +
...\label{equation}
\end{eqnarray}
The integrals $I^{(L)}(x,y)$ are given essentially by Hankel functions , viz
\begin{eqnarray}
&&I^{(1)}(x,y) = \int_{0}^{\infty}
  \frac{d{\alpha}}{{\alpha}}e^{-x\alpha-\frac{y}{\alpha}} =
  \frac{1}{2}\bigg[i{\pi}H_0^{(1)}(2i\sqrt{xy})
  +h.c. \bigg]\nonumber\\ 
&&~I^{(L)}(x,y) = \int_{0}^{\infty}
  \frac{d{\alpha}}{{\alpha}^L}e^{-x\alpha-\frac{y}{\alpha}} =
  \frac{1}{2}\bigg[\frac{i{\pi}}{L-1}(\frac{x}{y})^{\frac{L-1}{2}}
  \sqrt{xy}e^{\frac{iL{\pi}}{2}} \big[H_{L-2}^{(1)}(2i\sqrt{xy}) +
  H_L^{(1)}(2i\sqrt{xy})\big] + h.c. \bigg]~,~L>1.\nonumber
\end{eqnarray}
Hankel functions admit the series expansion $H_0^{(1)}(z)
=\frac{2i}{\pi} \ln z+...$ and $H_{\nu}^{(1)}(z) =
-\frac{i(\nu-1)!}{\pi}(\frac{2}{z})^{\nu}+..$ for $\nu>0$ when
$z{\longrightarrow}0$. In this case the mass $m$ and the external
momentum $E$ are both small compared to the cut-off $\Lambda=2/\theta$
and thus the dimensionless parameters $z{\equiv}\sqrt{xy} =
2\frac{m}{\Lambda}\frac{E}{\Lambda}$ or $z{\equiv}\sqrt{xy} = 2\sqrt{1
+ \frac{m^2}{{\Lambda}^2}}\frac{E}{\Lambda}$ are also small, in other
words we can calculate for example $I^{(1)}(x,y) = -2
\ln(2\sqrt{xy})$, $I^{(2)}(x,y) = 2x \ln(2\sqrt{xy}) + \frac{1}{y}$
and $I^{(L)}(x,y) = \frac{(L-2)!}{y^{L-1}}[1-\frac{xy}{(L-2)(L-1)}]$
for $L{\geq}3$. Thus the first term $N=0$ in the above sum ( i.e
euqation (\ref{equation})) is simply given by
\begin{eqnarray}
{\Delta}m^{NP}(p)=-\frac{m^2}{16{\pi}^2}ln(1+\frac{{\Lambda}^2}{m^2})+....
\end{eqnarray}
As one can see it does not depend on the external momentum $p$ at
all. In the commutative limit $\theta{\rightarrow}0$, this
diverges logarithmically as $ln{\Lambda}$ which is subleading compared
to the quadratic divergence of the planar function. Higher corrections
can also be computed and one finds essentially an expansion in
$\frac{{\Lambda}{\theta}^2E}{2}=E\theta=2\frac{E}{\Lambda}$ given by
\begin{eqnarray}
{\Delta}m^{NP}(p) &=& -\frac{m^2}{16{\pi}^2}
\ln(1+\frac{{\Lambda}^2}{m^2}) \nonumber\\
&+& \frac{{\Lambda}^2}{16{\pi}^2}I^{(1)}(x,y)
\sum_{p=2}^{\infty}\frac{1}{p!} \bigg(\frac{{\Lambda}{\theta}^2E}{2}
\bigg)^{2(p-1)}{\eta}_{p-1,p-2} + \frac{1}{16{\pi}^2}I^{(2)}(x,y)
\sum_{p=2}^{\infty}\frac{1}{p!} \bigg(\frac{{\Lambda}{\theta}^2E}{2}
\bigg)^{2p}{\eta}_{p,p-2} \nonumber\\
&+& \frac{1}{16{\pi}^2} \sum_{N=1}^{\infty}
\bigg(\frac{{\theta}^4E^2}{4} \bigg)^NI^{(N+2)}(x,y)
\sum_{p=2}^{\infty}\frac{1}{p!}
\bigg(\frac{{\Lambda}{\theta}^2E}{2}\bigg)^{2p}{\eta}_{p+N,p-2}.\nonumber 
\end{eqnarray}
[$x=m^2+{\Lambda}^2$, $y=\frac{{\theta}^4E^2}{4}$, ${\eta}_{p+N,p-2} =
  \sum_{M=0}^{p-2}\frac{(-1)^{M}}{M!(p+N-M)!}$]. It is not difficult
  to find that the leading terms in the limit of small external
  momenta (i.e. $E/\Lambda<<1$) are effectively given by
\begin{eqnarray}
{\Delta}m^{NP}(p) &=& -\frac{m^2}{16{\pi}^2} \ln\bigg(1 +
\frac{{\Lambda}^2}{m^2}\bigg) - \frac{E^2}{4{\pi}^2}
\ln\bigg(4\frac{E}{\Lambda}\sqrt{1 + \frac{m^2}{{\Lambda}^2}}\bigg)
\bigg[1 + O\bigg(\frac{E^2}{{\Lambda}^2}\bigg)\bigg] +
\frac{E^2}{8{\pi}^2} \bigg[1 +
  O^{'}\bigg(\frac{E^2}{{\Lambda}^2}\bigg)\bigg]. \label{np1}\nonumber\\
\end{eqnarray}
Clearly in the strict limit of small external momenta when
$E{\rightarrow}0$, we have $E^2 \ln E{\rightarrow}0$ and the
non-planar contribution does not diverge (only the first term in
(\ref{np1}) survives this limit as it is independent of $E$) and hence
there is no UV-IR mixing. The limit of zero noncommutativity is
singular but now this divergence has the nice interpretation of being
the divergence recovered in the non-planar $2-$point function when the
cut-off ${\Lambda} = \frac{2}{\theta}$ is removed. This divergence is
however logarithmic and therefore is sub-leading compared to the
quadratic divergence in the planar part.

The effective action (\ref{effective2}) with $j=[2\sqrt{l}]$ can be
obviously obtained from quantizing the action (\ref{ncr4action}) with
the replacements ${\mu}_l^2{\rightarrow}m^2$,
${\phi}_3{\rightarrow}{\phi}_2{\equiv}{\phi}_2(X^{NC}) = \int
\frac{d^4p}{(2{\pi})^4} {\phi}_{2} (\vec{p}) e^{-ipX^{NC}} =
{\phi}_2^{\dagger}$ and where as before we have to regularize all
integrals in the quantum theory with a cut-off $\Lambda=2/\theta$. The
star product $*$ is the Moyal-Weyl product given in (\ref{star}) with
the substitutions ${\theta}'{\rightarrow}{\theta}$,
$x^{NC}{\longrightarrow}X^{NC}$. This effective action can also be
rewritten in the form
\begin{equation} 
\int d^4x \bigg[\frac{1}{2} {\partial}_{\mu}{\phi}_2*_{\Lambda}
  {\partial}_{\mu}{\phi}_2 + \frac{1}{2} m^2{\phi}_2*_{\Lambda}
  {\phi}_2 + \frac{g_4^2}{4!}  {\phi}_2* {\phi}_2*
  {\phi}_2*{\phi}_2\bigg],
\label{action1}
\end{equation} 
which is motivated by the fact that the {\it effective} star product
defined by
\begin{eqnarray}
f*_{\Lambda}g(X^{NC}) &=& \int_{{\Lambda}} \frac{d^4p}{(2{\pi})^4}
f(\vec{p})\int_{{\Lambda}} \frac{d^4k}{(2{\pi})^4} g(\vec{k})
e^{-ipX^{NC}}*e^{-ikX^{NC}} \nonumber\\
&=& \int d^4y'd^4z' {\delta}^4_{\Lambda}(y') {\delta}^4_{\Lambda}(z')
f(y-y')*g(z-z')|_{y=z=X^{NC}},
\label{prod}
\end{eqnarray}
is such that $\int d^4xf*_{\Lambda} g(x) = \int_{{\Lambda}}
\frac{d^4p} {(2{\pi})^4} f(\vec{p}) g(-\vec{p})$. The distribution
${\delta}^4_{\Lambda}(y')$ is not the Dirac delta function
${\delta}^4(y')$ but rather ${\delta}^4_{\Lambda}(y') =
\int_{{\Lambda}} \frac{d^4p} {(2{\pi})^4} e^{-ipy'}$,
i.e. ${\delta}^4_{\Lambda}(y')$ tends to the ordinary delta function
in the limit ${\Lambda}{\rightarrow}{\infty}$ of the commutative plane
where the above product (\ref{prod}) also reduces to the ordinary
point-wise multiplication of functions. If the cut-off ${\Lambda}$ was
not correlated with the non-commutativity parameter ${\theta}$, then
the limit ${\Lambda}{\rightarrow}{\infty}$ would had corresponded to
the limit where the product (\ref{prod}) reduces to the Moyal-Weyl
product given in equation (\ref{star}). This way of writing the
effective action (i.e. (\ref{action1})) is to insist on the fact that
all integrals are regularized with a cut-off ${\Lambda}=2/\theta$. In
other words the above new star product which appears only in the
kinetic part of the action is completely equivalent to a sharp cut-off
${\Lambda}$ and yields therefore exactly the propagator (\ref{presc})
with which only modes ${\leq}{\Lambda}$ can propagate.

We should also remark here regarding non-locality of the star product
(\ref{prod}). At first sight it seems that this non-locality is more
severe in (\ref{prod}) than in (\ref{star}), but as it turns out this
is not entirely true: in fact the absence of the UV-IR mixing in this
product also suggests this. In order to see this more explicitly we
first rewrite (\ref{prod}) in the form
\begin{eqnarray}
&& f*_{{\Lambda}}g(X^{NC}) = \int d^4y' d^4z' f(y') g(z')
  K_{\Lambda}(y',z';X^{NC}) \nonumber\\
&& K_{\Lambda}(y',z';X^{NC}) = {\delta}^4_{\Lambda}(y-y')*
  {\delta}^4_{\Lambda}(z-z')|_{y=z=X^{NC}}. \nonumber
\end{eqnarray}
The kernel $K_{\Lambda}$ can be computed explicitly and is given
by
\begin{equation} 
K_{\Lambda}(y',z';X^{NC}) = \int_{{\Lambda}} \frac{d^4k}{(2{\pi})^4}
{\delta}^4_{\Lambda}(X^{NC}-y' + \frac{{\theta}^2}{2}Bk) e^{ik(z' -
  X^{NC})}. \nonumber
\end{equation} 
For the moment, let us say that $\Lambda$ and $\theta$ are
unrelated. Then, taking ${\Lambda}$ to infinity gives \cite{mirase,dn}
\begin{equation} 
K(y',z';X^{NC}) = \frac{16}{{\theta}^8 \det B} \frac{1}{(2{\pi})^4}
e^{\frac{2i}{{\theta}^2}(z' - X^{NC})B^{-1}(y' - X^{NC})}. \nonumber
\end{equation} 
If we have for example two functions $f$ and $g$ given by
$f(x)={\delta}^4(x-p)$ and $g(x)={\delta}^4(x-p)$, i.e. they are
non-zero only at one point $p$ in space-time, their star product which
is clearly given by the kernel $K(p,p;X^{NC})$ is non-zero everywhere
in space-time. The fact that $K$ is essentially a phase is the source
of the non-locality of (\ref{star}) which leads to the UV-IR mixing.

On the other hand the kernel $K_{\Lambda}(p,p;X^{NC})$ with finite
${\Lambda}$ can be found in two dimensions (say) to be given by
\begin{equation} 
K_{\Lambda}(p,p;X^{NC}) = \frac{1}{{\pi}^2{\theta}^4}
\int_{-{\theta}}^{{\theta}} da{\delta}_{\Lambda}(a +
L_1) e^{\frac{2i}{{\theta}^2}L_2a}
\int_{-{\theta}}^{{\theta}} db{\delta}_{\Lambda}(b +
L_2) e^{-\frac{2i}{{\theta}^2}L_1b},\nonumber
\label{effestar}
\end{equation} 
with $L_a=X_a^{NC}-p_a$, $a=1,2$. If we now make the approximation to
drop the remaining ${\Lambda}$ (since the effects of this cut-off were
already taken anyway) one can see that the above integral is non-zero
only for $-{\theta}+p_1{\leq}X_1^{NC}{\leq}{\theta}+p_1$ and
$-{\theta}+p_2{\leq}X_2^{NC}{\leq}{\theta}+p_2$ simultaneously. In
other words the star product $K_{\Lambda}(p,p;X^{NC})$ of $f(x)$ and
$g(x)$ is also localized around $p$ within an error ${\theta}$ and is
equal to $\frac{1}{{\pi}^2{\theta}^4}$ there . The star product
(\ref{prod}) is therefore effectively local.

Final remarks are in order. First we note that the effective star
product (\ref{prod}) leads to an effective commutation relations
(\ref{commutationrelation}) in which the parameter ${\theta}^2$ is
multiplied by an overall constant equal to $\int d^4y' d^4z'
{\delta}^4_{\Lambda}(y'){\delta}^4_{\Lambda}(z')$, we simply skip the
elementary proof. Remark also that this effective star product is
non-associative as one should expect since it is for all practical
purposes equivalent to a non-trivial sharp momentum cut-off ${\Lambda}$
.

The last remark is to note that the prescription (\ref{presc}) can
also be applied to the canonical limit of large stereographic
projection of the spheres onto planes, and in this case one can also
obtain a cut-off ${\Lambda}' = \frac{2}{{\theta}'}$ with $j$ fixed as
above such that $j = [2\sqrt{l}]$. The noncommutative plane
(\ref{comm1}) defined in this way is therefore completely equivalent
to the above noncommutative plane (\ref{commutationrelation}).

{\bf Acknowledgments} It is a pleasure to thank Denjoe O'Connor
and Peter Presnajder for discussions. SV would like to thank DIAS
for warm hospitality during the final stage of this project. The
work of SV is supported in part by DOE grant DE-FG03-91ER40674.

\end{document}